\documentclass[useAMS,usenatbib]{mn2e}
\usepackage{graphicx}

\title[Tidal radii of globular clusters and the mass of the Milky Way]
{Tidal radii of globular clusters and the mass of the Milky Way}
\author[M. Bellazzini]{M. Bellazzini$^{1}$\thanks{E-mail:
bellazzini@bo.astro.it}\\
$^{1}$INAF - Osservatorio Astronomico di Bologna, via Ranzani 1, 40127, Bologna,
Italy}
\begin{document}

\date{Accepted for publication by MNRAS}

\pagerange{\pageref{firstpage}--\pageref{lastpage}} \pubyear{2003}

\maketitle

\label{firstpage}

\begin{abstract}
Tidal radii of remote globular clusters ($R_{GC}\ge 35$ kpc) are used to provide
constraints of the mass profile of the Milky Way galaxy that are independent
of kinematic data. The available data are consistent with the profile of an
isothermal sphere with circular velocity $V_c = 220 \pm 40$ km/s in the radial
range $35$ kpc $\le R_{GC}\le 100$ kpc, in good agreement with all recent
estimates. The more robust constraint at large distances from the galactic
center is provided by NGC~2419, yelding an enclosed mass of
$1.3^{+2.9}_{-1.0}\times 10^{12} ~M_{\sun}$ at $R_{GC}\simeq 90$ kpc. 
\end{abstract}


\section{Introduction}

One of the most straightforward examples of the great difficulties associated
with the measure of basic physical quantities on the astrophysical scales is
provided by the quest for the mass of the Milky Way. Even applying the most 
refined analysis and using the whole wealth of available data,
realistic uncertainties affecting single estimates of such a fundamental 
parameter tipycally amount to $\sim 100 - 300$ \% 
\cite[see, e.g.][for a state-of-the-art analysis]{we99}.
The whole problem was recently reviewed and critically discussed by
\citet[][hereafter Z99]{z99}. This author notes that the concept of total mass
of the Milky Way is somehow ill-defined since we ignore the actual extent of the
Dark Matter (DM) halo of the Galaxy. Hence, it is much safer to refer mass
estimates to the {\em enclosed mass} within the galactocentric distance
($R_{GC}$) sampled by the mass-tracer under consideration. 
This approach allows a
sensible comparison between different estimates, since consistency requires that
all estimates shall be in agreement with a unique mass profile [$M(R_{GC})$]
over the whole range of galactocentric distances that can be probed. 
Z99 uses the {\em isothermal sphere} as a reference model to perform such a
comparison and concludes that all the available estimates, covering different
ranges in $R_{GC}$ and using different tracers (from the HI rotation curve to
the outermost Galactic satellites), are consistent with the mass profile of
an isothermal sphere
with rotational velocity $V_c \sim 180$ km/s. Note that in this context the
isothermal sphere is (obviously) {\em not} intended as a realistic model
for the Galactic DM halo, but just as a suitable mass profile to provide
comparison and cross-validation of the various estimates. Hence, despite the
large uncertainties affecting the single mass estimates, a remarkable overall 
consistency is apparent. If interpreted in the classical Newtonian framework,
the mass of the Galaxy ($M_G$) is observed to grow approximately as 
$M_G\propto R_{GC}$ and the enclosed mass within $R_{GC}\simeq 300$ kpc is of 
order $M_G \sim 2\times 10^{12} ~M_{\sun}$ (Z99).

All the estimates considered by Z99 (as well as the more recent ones in
the post-'99 literature) rely on the kinematics of the adopted tracers
(rotational velocity curve, escape velocity of local stars, motion of
satellites, either stars, globular clusters or dwarf galaxies). Therefore, an
estimate of $M_G(R_{GC})$ not based on kinematical data would provide a further
important consistency check of our ideas on the mass profile of the Galaxy.
Such kind of probe may be provided by tidal radii of globular clusters (GC) 
\citep{vh,k62}. Theory predicts that the Galactic tidal field fixes the cut-off 
in the density profile of GCs, the position of the cut-off depending on the
cluster distance and on the ratio between the mass of the Galaxy and the mass
of the cluster ($m_c$). Hence, having $m_c$ and $R_{GC}$ from observations, an
estimate of $M_G$ may be obtained, though affected by large uncertainties.
The approach has been attempted in the past by \citet[][hereafter W81]{waka} and
by \citet[][hereafter IHW]{ihw}. W81 used globular clusters and 
dwarf spheroidal galaxies as tracers, and obtained 
$M_G(78~kpc)=2.0^{+1.3}_{-0.7} \times 10^{11} ~M_{\sun}$.
The result is in marginal disagreement with the conclusions by Z99 but is
probably affected by the large uncertainties in the distance and tidal radii of
dwarf spheroidals (for the same reasons these galaxies are not considered in the
present analysis). 
IHW used only GCs and found $M_G(44~kpc)=8.9\pm 2.6 \times 10^{11} ~M_{\sun}$,
consistent, within the uncertainties, with the Z99 results.

There are cogent reasons to try to repeat the experiment twenty years later.
The quality of the observational material is greatly improved, mainly thanks to
the extensive compilation of surface brightness profiles by \citet{tdk95}, and
theoretical advancements put the problem in a new light 
\citep{ola92,ola95,meziane,bog99}. Moreover the impressive growth of computing
facilities allow an accurate analysis of the uncertainties (a
particularly critical point for this kind of application, see IHW for
a discussion) by means of extensive Montecarlo simulations. 

In this paper I review the use of tidal radii of globular clusters in the light
of the more recent theoretical and observational result and I check if the mass
estimates from this technique are consistent with ``kinematical'' estimates,
dealing in particular with the mass profile at large $R_{GC}$, e.g. the most
interesting range (Sect.~2). The 
main conclusions and future prospects are summarized in Sect.~3.

\section[]{Tidal radii as mass probes}

A general formula for the tidal radius ($r_t$) of a globular cluster of mass 
$m_c$ orbiting around a galaxy of mass $M_G(R)$ is (\citealt{k62,ola92}, 
IHW, W99):

\begin{equation}
r_t=k\left[\frac{m_c}{f(e)M_G(R_p)}\right]^{\frac{1}{3}}R_p
\end{equation}

where $R_p$ is the peri-galactic distance, $f(e)$ is a function depending on
the form of the galactic potential and on the eccentricity of the cluster orbit
$e$ and $k$ is a factor, firstly introduced by \citet{ke81a,ke81b} to account 
for the elongation of the limiting tidal surface along the line between the 
cluster center and the galactic center. According to this author, IHW and W99
assume $k=\frac{2}{3}$. From now on we drop, for brevity, the explicit
indication of the dependence of mass on distance: in any case $M_G$ shall be
intended as the mass enclosed with a given galactocentric distance $R$ (the
correponding index is also dropped, from now on $R$ means $R_{GC}$).

\begin{figure}
\includegraphics[width=84mm]{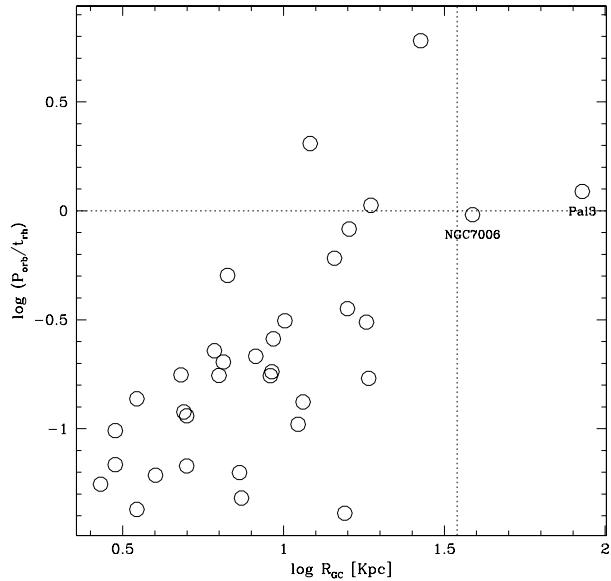} 
\caption{The logarithm of the ratio between the orbital period (P) and the
half-mass relaxation time is plotted versus the galactocentric distance for all
the Galactic globulars for which orbital parameter are available in
\citet{dana1} or \citet{dana2}. 
The horizontal line marks the threshold $P/t_{rh}=1$, the
vertical line marks $R_{GC} = 35$ Kpc. The names of the clusters 
with $R_{GC} > 35$ Kpc are also reported.}
\end{figure}

The dependence on $e$ was introduced by \citet{k62} by reformulating the
equation for the instantaneous tidal radius of \citet{vh} in terms of the
perigalactic distance (instead of the instantaneous distance $R$). This choice
was motivated by the fact that the typical orbital period ($P$) of globulars was
expected to be much shorter than the internal relaxation time (tipically
quantified by the half-mass relaxation time $t_{rh}$). 
In this case 
two-body relaxation is unable to keep the external structure of the cluster at
pace with the changing galactic tidal field. In other words, the tidal force at
perigalacticon truncates the cluster at the corresponding radius and the
internal relaxation is too slow to restore a larger limiting radius before the
next perigalactic passage \cite[see also][]{ola95}.

However recent comparisons between predicted and observed tidal radii that
included all the information on cluster orbits \cite[as derived from measured
proper motions][]{meziane,Od97,bog99} showed that observed tidal radii are
larger than one would expect if they were fixed at the perigalactic point.
\citet{meziane} suggest that the actual tidal radius depends on the orbital
phase, while \citet{bog99} argue that a suitable average along the orbit
defines a much more proper tidal radius with respect to the perigalactic value.

\citet{ola92} and \citet{ola95} studied in detail the properties of tidal radii
of globular clusters by means of N-body simulations. They found that while
the tidal limit of clusters with $P/t_{rh} << 1$ is effectively set at
perigalacticon, in clusters with $P/t_{rh} \sim 1$ internal relaxation effects
are able to repopulate their external regions after the perigalacticon passage
and their tidal radius may be comparable to the value expected at 
{\em apogalacticon}. 

In Fig.~1 the logarithm of the $Q_t=P/t_{rh}$ ratio is plotted versus the
galactocentric distance for all the clusters for which orbital parameters from
measured proper motions are available 
\cite[from the computations by][]{dana1,dana2}. 
The relaxation
times are from the 2003 version of the \citet{h96} catalogue, which is
the source of all the globular cluster data used in this study if
not otherwise stated. There is a clear correlation between $Q_t$ and $R$
suggesting that beyond $R\sim 35$ kpc clusters are very likely to have 
$Q_t\ga 1$, i.e., according to the results by \citet{ola95}, 
their $r_t$ does not reflect the tidal force at perigalacticon.

On the basis of all the above considerations I make the assumption that {\em the
observed tidal radii of distant globular clusters (e.g., those with $R\ge 35$
kpc) are probes of the Galactic tidal force at their present
position}. In other words I take the galactocentric distance 
$R_p\le R \le R_a$ as a reasonable approximation of the effective orbital radius
at which the tidal radius of clusters with $Q_t\ga 1$ is fixed 
\citep{meziane,mh97,Od97,bog99,ola95}. The assumption limits the range that can
be probed to the outer halo of the Galaxy. On the other hand this is the most
interesting range since the contribution of barionic matter to the mass budget
is expected to be negligible in this region. 
Furthermore, clusters orbiting in such remote regions
are less likely to have their structure changed by close encounters with the
galactic disk, bulge or with large molecular clouds, and are therefore
more reliable probes of the Galactic tidal field \cite[see][for an extensive
review of the mechanisms affecting the structure of GCs]{mh97}.

Following the above assumption, the following form of the theoretical
tidal radius is adopted:

\begin{equation}
r_t=k\left[\frac{m_c}{2M_G}\right]^{\frac{1}{3}}R 
\end{equation}
 
e.g, Eq.~1 with $f(e)=2$ and $R$ instead of $R_p$. In the context of a
logarithmic potential this is the exact formula for the tidal radius of a
cluster on a circular orbit \citep{k62}.
Solving with respect to
$M_G$:

\begin{equation}
M_G = \frac{1}{2} k^3  m_c \left(\frac{R}{r_t}\right)^3 ~.
\end{equation}
 
W81, IHW and \citet{ola95} derived $f(e)$ for eccentric orbits in a logarithmic 
potential.
It is important to note that the adoption of $f(e)=f_{log}(e)$, instead of the
instantaneous value $f(e)=f_{log}(e=0)$ adopted here, has a modest 
impact on the final $M_G$ estimates. 
The adoption of the $f_{log}(e)$ by \citet{ola95} changes
the final $M_G$ estimate by a factor $\le 2$ for $e\le 0.825$ and by a factor 
$\le 4$ for $e\le 0.935$ with respect to the $f(e)=2$ case adopted here. 
In any case
the derived $M_G$ is larger than what obtained with Eq.~3. In analogy with the
analysis that will be described in Sect.~2.2 I performed an extensive set of
Montecarlo experiments using the $f_{log}(e)$ version of Eq.~1 and exploring 
the whole range of possible orbital eccentricities. 
The resulting mass (and mass
profile) estimates are fully consistent with the results obtained from Eq.~3.
Hence, the approximations involved in the adopted approach appear fully 
adequate for the present purpose \cite[see also][]{rod_sphe}.

\subsection{The observational side}

Tidal radii of GCs are not, in general, directely observable quantities. Fitting
King models \citep{k62} to surface brightess profile of globular clusters one
obtains an estimate of the core radius ($r_c$) and of the concentration
parameter $C =$ log $\left(\frac{r_t}{r_c}\right)$; $r_t$ is derived from these
fitted parameters. 
Hence, the best estimates of $r_t$ may be obtained from 
bright clusters (e.g., providing high signal-to-noise data for the fit) 
whose surface brightness profile is reliably measured over the largest possible
radial range (thus limiting at a minimum the extrapolation to the actual tidal
limit). 

Unfortunately the range $R_{GC}>35$ kpc is mainly populated by sparse,
low luminosity clusters whose tidal radii and integrated magnitudes
are quite uncertain. 
The suitable galactic clusters in the relevant range of galactocentric 
distances are:
Pal~15, NGC~7006, Pyxis, Pal~14, NGC~2419, Eridanus, Pal~3, Pal~4 and AM-1.
Pal~15 and Pyxis have been excluded since they are affected by significant
amount of interstellar extinction ($A_V>0.6$ mag, that implies a larger
uncertainty in the estimate of $m_c$). Note however that the inclusion of these
clusters does not change in any way the final results presented below. 
Of the remaining clusters the best suited for the present analysis are NGC~7006
and NGC~2419, two bright and well studied clusters. In particular NGC~2419 is
the 4-th most luminous cluster of the whole Galaxy ($M_V=-9.58$) and its surface
brightness profile have been reliably measured out to the 85 \% of the deduced
tidal radius, i.e. the level of extrapolation is quite modest. Finally NGC~2419
is the only cluster of the considered set for which a direct estimate of the
mass-to-light ratio ($M/L$, a fundamental ingredient to derive $m_c$) is
available \cite[$M/L_V=1.2$, from][]{pm93}. 

In conclusion, NGC~7006 and NGC~2419 provide by far the most robust and less
uncertain mass probes, the remaining clusters are retained just for consistency
check.

\subsubsection{Montecarlo simulations}

$M_G$ is estimated from each cluster using Eq.~3. To deal with uncertainties I
obtained 10000 independent $M_G$ estimates for each cluster by extracting at 
random (from suitable distributions described below) the following quantities:
the observed V-band distance modulus $\mu_V$, the apparent integrated V
magnitude $V_t$, the tidal radius in arcmin $r_t$, 
the V-band mass-to-light ratio in
solar units $M/L$ and Keenan's $k$ factor (see below). The color excess
[$E(B-V)$] is kept fixed, since it is small (less than $\sim 0.1$ mag) 
in all of the considered cases.
For each set of extracted parameter the following items are computed: the
distance from the Sun and the Galactic Center from $\mu_V$, $E(B-V)$ and the
galactic coordinates, the linear tidal radius from its angular value and
distance, the mass of the cluster from $V_t$, $\mu_V$ and $M/L$, and finally,
from $R$ and $r_t$ in kpc and $m_c$ in solar masses, the Galactic mass enclosed
within $R$.  
All the distributions are
chosen to (conservatively) cover the whole range that is
compatible with the adopted uncertainties of each parameter, 
hence the final 10000 $M_G$ estimates
cover the whole range allowed by taking into account {\em all the possible
sources of error}. Finally the median of the 10000 estimates is computed as well
as the range in $M_G$ including the 90 \% of the derived $M_G$ estimates.

\subsubsection{Uncertainties on input parameters.}

To have a closer look to the details of the simulations, 
the assumptions on input parameters are shortly described below.

\begin{itemize} 

\item $k$ factor. Keenan's factor ($k=\frac{2}{3}$) provides an average 
correction for the non-spherical shape of the limiting tidal surface. To account
for the possible cluster-to-cluster variation of this parameter, $k$ is
extracted from a uniform distribution in the range $0.5\le k \le 1.0$
\cite[see also][]{heggie}.

\item $\mu_V$. To account for both the measurement errors and the uncertainties
still affecting the distance scale of globular clusters \cite[see][and
references therein]{carla} 
$\mu_V$ values were extracted from a gaussian distribution with mean equal to
the $\mu_V$ listed by \citet{h96} and standard deviation $\sigma_\mu=0.1$ mag 
for NGC~2419 and NGC~7006 and $\sigma_\mu=0.2$ mag for the remaining, 
less extensively studied, clusters.

\item $V_t$ values were extracted from a gaussian distribution with mean equal 
to the $V_t$ listed by \citet{h96} and standard deviation $\sigma_V=0.1$ mag 
for NGC~2419 and NGC~7006 and $\sigma_V=0.2$ mag for the remaining clusters.
Note that the measure of this parameter is particularly critical for low-density
and low-brightness clusters, like the majority of those considered here.

\item $M/L$. The observed Fundamental Plane of GCs \citep{d95,mfp} implies 
that the $M/L$ ratio of globular clusters is constant to within a factor of a
few, approximately compatible with the measurement errors. \citet{pm93} find 
$0.5 \sol M/L \sol 4$ in good agreement with \citet{mandu}. Here the $M/L$
values are extracted from a gaussian distribution with mean  $M/L=1.2$ (e.g.,
the estimated $M/L$ of NGC~2419 and the average $M/L$ derived by
\citealt{mandu}) and $\sigma_{M/L}=0.4$, with the further constraint 
$M/L\ge 0.5$. This assumption ensures that the range of observed $M/L$ of GCs is
fully explored.

\item $r_t$. I searched the literature to find estimates of $r_t$ for the
considered clusters that may supersede those reported by \citet{h96}, all drawn
from \citet{tdk95}. The only (partially) successful case was Pal~14, 
for which I
adopt the estimate by \citet{hp14} whose density profile is marginally more
extended with respect to that by \citet{tdk95}. The adopted $r_t$ values are
extracted from a gaussian distribution with the mean equal to the listed values 
and with standard deviation in the range $\sigma_r= 0.1 r_t - 0.3 r_t$, 
depending  on the quality and extension of the available surface brightness 
profile. In particular $\sigma_r= 0.1 r_t$ for NGC~2419, $\sigma_r= 0.2 r_t$ 
for NGC~7006 and $\sigma_r= 0.3 r_t$ for the remaining clusters.

\end{itemize}

The Referee correctly pointed out that the estimates of tidal radii may be
also plagued by systematics. For example the adoption of different models
\cite[e.g.][]{wilson} to fit the surface brightness profiles of 
globulars may lead to obtain significantly larger limiting radii than what
estimated with King's models \cite[see][for an application]{meylan}. 
Moreover, some theoretical studies also suggest that realistic models of 
globular clusters in the Galactic tidal field may be slightly more spatially 
extended than King's models \cite[see, e.g.][]{kash,heggie}.
To explore
the effect of systematics that may change the observationally estimated $r_t$ 
values up to a factor $\sim 2$ I repeated the analysis described above for NGC~2419
\cite[$rt=8.74\arcmin$, according to][]{tdk95} assuming $\sigma_r= 0.5 r_t$
instead of $\sigma_r= 0.1 r_t$. To preserve compatibility with the observed
profile, which reaches $r\simeq 7.5\arcmin$, I forced the randomly extracted
tidal radii to the range $r_t>6\arcmin$. Hence the final $r_t$ of the Montecarlo
simulation are in the range  $6\arcmin<r_t\sol 20\arcmin$. Note that at 
$r\ge 6\arcmin$
the observed profile is rapidly falling and the surface brightness is
$\Sigma_V>28 $ mag/arcsec$^2$, thus it is quite unlikely that the actual
limiting radius is much larger than 10$\arcmin$. With these new assumptions the
typical uncertainty on the final $M_G$ estimates grows from $\sim 75$\% 
(for the case $\sigma_r= 0.1 r_t$) to $\sim 110$\% while the derived median
$M_G$ is practically unchanged. Therefore the inclusion of these possible
systematic errors in the uncertainty budged doesn't seriously affect the main
conclusions of this paper (see Sect.~2.2), at least for what concern the cluster
that provide the most interesting constraint on $M_G$, e.g. NGC~2419.

\begin{figure*}
 \includegraphics[width=168mm]{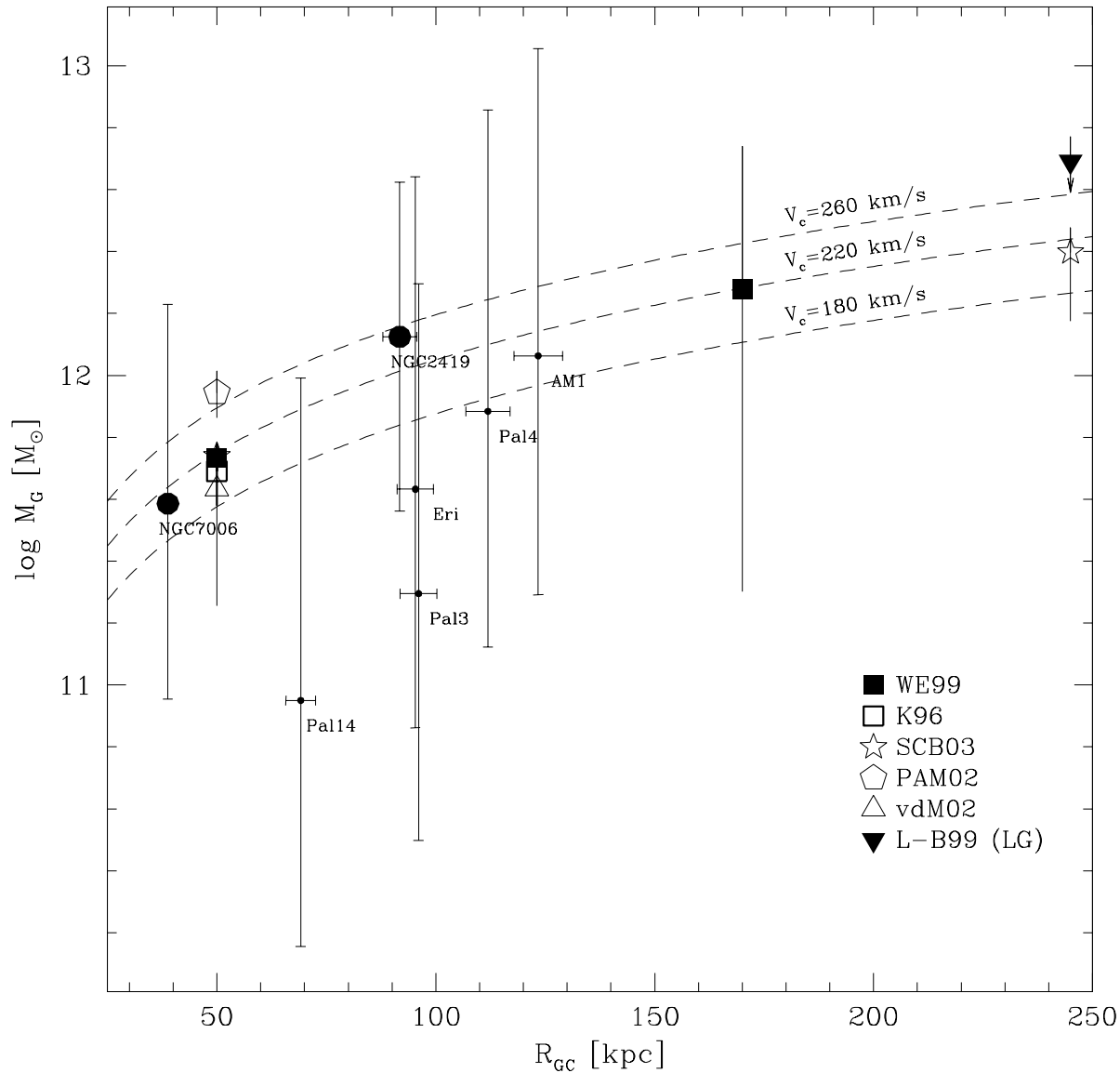} 
 \caption{The Mass profile of the Milky Way halo (for $R_{GC}>35$ kpc) from 
  various recent estimates based on kinematical data (see legend in the lower 
  left corner) and
  from the tidal radii of the selected globular clusters (filled circles). 
  Large
  filled circles are the best observed globulars, providing (by far) the 
  stronger constraint on $M_G$ with respect to the other clusters. The error
  bars for globulars enclose the  90 \% c.l. range.
  The mass profiles of isothermal spheres with $V_c=180,220,260$ km/s 
  are also reported for reference. Acronyms: WE99 = \citet{we99}; K96 =
  \citet{k96}; SCB03 = \citet{scb}; PAM02 = \citet{pam}; vdM03 = \citet{vdm};
  L-B99 = \citet{lb99}, mass of the Local Group. The arrow indicates that the
  last one is an upper limit for the total mass of the Galaxy.}
\end{figure*}

\subsection{The mass of the Galaxy within $35<R_{GC}<100$ kpc.}

Fig.~2 reports the median $M_G$ and $R$ of the 10000 simulation carried on for
each considered cluster (filled circles). The error bars enclose 90 \% of the
derived estimates ($\pm 45$ \%). The most reliable points (NGC~7006 and
NGC~2419) are indicated by larger symbols.

Other mass estimates, more recent than those considered by Z99, are also
reported (see legend). The only pre-99 estimate reported is that by \citet{k96},
who provides the more extensive treatment of the problem at that epoch.
The filled triangle with an arrow, labeled L-B99 in the legend, provides a
sensible upper limit to the total mass of the Galaxy since it is the estimate of
the mass of the Local Group obtained by \citet{lb99} using the Local Group
timing technique.
The reported error bars have etherogeneous meanings 
($1-\sigma$ errors, 90 \% c.l., etc., as provided by the authors) and hence 
they are not directly comparable.
The mass profiles of isothermal spheres with 
$V_c=180,220,260$ km/s are also plotted for reference. They are intended to
allow the comparison of the estimates at large R 
with constraints provided by the rotational velocity of the HI disk at
$R<20$ kpc, taking into account the whole range of possible uncertainty on the
Galactic $V_c$ \cite[see][Z99, and references therein]{ft91,pam}.

From the inspection of Fig~2 the following main conclusions can be drawn:

\begin{enumerate}

\item All the reported estimates based on kinematical data are in agreement,
      within the uncertainties, with the reported mass profiles and with the
      results summarized by Z99.
      
\item All the estimates based on the tidal radii of globular clusters are
      consistent with the reported mass profile, within the (large)
      uncertainties.
      
\item The clusters that provide the most reliable constraint with the adopted
      technique (e.g., NGC~7006 and NGC~2419) give estimates of the enclosed
      Galactic mass in {\em excellent} agreement with the reported profiles and
      with all other estimates. In particular, from NGC~7006 I obtain 
      $M_G($39 kpc$)=0.38^{+1.3}_{-0.1}\times 10^{12} ~M_{\sun}$, and from 
      NGC~2419, $M_G($92 kpc$)=1.3^{+2.9}_{-1.0}\times 10^{12} ~M_{\sun}$.
      
\end{enumerate}

Hence, the present application of the tidal radii technique provide an 
independent
validation of the {\em standard} framework for the mass of the Milky Way as it
emerges from the analysis by Z99 and from more recent studies. There is a quite
remarkable general agreement among all the considered estimates, indicating
that (a) the mass of the Galaxy appears to grow with galactocentric distance at
least up to $R\ga100$ Kpc and (b) the total mass of the Galaxy is larger than
$\sim 2\times 10^{12} ~M_{\sun}$. This implies that the V-band mass to light 
ratio of the Milky Way spans a range $20 \sol M/L \sol 100$ in the radial range
$30$ kpc $ \sol R \sol 200$ kpc (see Fig.~3 by Z99). Hence, in the classical
Newtonian theory, all the available observational constraints consistently 
point to the conclusion that the Milky Way is surrounded by a huge 
Dark Matter halo.

\section{Conclusions}

The tidal radii of galactic globular clusters have been used to obtain estimates
of the enclosed mass of the Galaxy in the range of galactocentric distances 
$35$ kpc $\le R_{GC}\le 100$ kpc, under the assumption that, for such remote
clusters, $R_{GC}$ is a reasonable approximation of the orbital radius at which
their tidal limit is imposed  \citep{meziane,Od97,bog99,ola95}.
The adopted technique provides an estimate of the enclosed galactic mass that is
independent of kinematical data that, on the other hand, provides the basis of
all other existing estimates (see Sect.~1). Therefore, while the associated
uncertainties are still quite large, the present application provide 
(at least) an
interesting consistency check of the existing kinematical estimates of $M_G$.

The estimates obtained from tidal radii are fully consistent with results from
other authors and other methods. All the available constraints are consistent
with the mass profile of an isothermal sphere with $V_c=220\pm40$ km/s. Hence,
the present analysis provides independent support to the fact that the mass of 
the Galaxy grows with $R$ out to large distances from the Galactic Center and 
that the mass enclosed within $R\simeq90$ kpc is 
$M_G \simeq 10^{12} ~M_{\sun}$.

The present analysis indicates that tidal radii of remote globular clusters
may be better probes
of the galactic potential than previously believed (IHW). 
A detailed  and comprehensive theoretical
analysis it is now at hand with realistic
N-body simulations \cite[e.g., with GRAPE,][]{makino} and can provide much
sounder and solid basis to the technique. At the same time, wide-field cameras
mounted on large telescopes may provide the opportunity to obtain more extended
density profiles, based on star counts, hence reducing the observational
uncertainty. These advancements may ultimately lead to a fully reliable
additional technique to probe the mass profile of our Galaxy, nicely 
independent of and complementary to the usual kinematical methods.

\section*{Acknowledgments}

I am grateful to L. Ciotti for many useful discussions.
L. Ciotti, R. Sancisi and F. Fusi Pecci are also acknowledged for a critical
reading of the original manuscript. The Referee is also warmly acknowledged for
a detailed analysis of the original manuscript as well as for many useful
suggestions.
This research is partially supported by
the italian {Ministero  dell'Universit\'a e della Ricerca Scientifica}
(MURST) through the COFIN grant p.  2001028879, assigned to the project {\em
Origin and Evolution of Stellar Populations in the Galactic Spheroid}.
The support of ASI is also acknowledged.
This research has made use of NASA's Astrophysics Data System Abstract
Service.

\label{lastpage}

\end{document}